\documentclass[apjl]{emulateapj}

\newcommand{\gray}{$\gamma$-ray}
\newcommand{\grays}{$\gamma$-rays}

\newcommand{\pubjournal}[5]{#4, #1, #2, #3}

\newcommand{\pubjournalb}[5]{#4, #5 #1 #3}

\shorttitle{Attenuation of VHE gamma rays}
\shortauthors{Moskalenko et al.}
\journalinfo{The Astrophysical Journal}
\submitted{Submitted 2005 November 2; accepted 2006 February 17}

\begin{document}

\title{Attenuation of VHE gamma rays by the Milky Way Interstellar Radiation Field}

\author{Igor V. Moskalenko\altaffilmark{1}}
\affil{
   Hansen Experimental Physics Laboratory, 
   Stanford University, Stanford, CA 94305
\email{imos@stanford.edu}}
\altaffiltext{1}{Also Kavli Institute for Particle Astrophysics and Cosmology,
Stanford University, Stanford, CA 94309}

\author{Troy A. Porter}
\affil{
  Department of Physics and Astronomy, Louisiana State University, 
  Baton Rouge, LA 70803
\email{tporter@lsu.edu}}

\and

\author{Andrew W. Strong}
\affil{Max-Planck-Institut f\"ur extraterrestrische Physik,
Postfach 1312, D-85741 Garching, Germany
\email{aws@mpe.mpg.de}}

\begin{abstract}
The attenuation of very high energy \grays\ by pair production 
on the Galactic interstellar radiation field has long been thought of as 
negligible. However, a new calculation of the interstellar radiation field 
consistent with multi-wavelength 
observations by DIRBE and FIRAS indicates that the energy density of the 
Galactic interstellar radiation field is higher, particularly in the Galactic
center, than previously thought. 
We have made a calculation of the 
attenuation of very high energy \grays\ in the Galaxy using this 
new interstellar radiation field which
takes into account its nonuniform spatial and angular 
distributions. 
We find that the maximum attenuation occurs around 100 TeV at the 
level of about 25\% for sources located at the Galactic center,
and is important for both Galactic and extragalactic sources.
\end{abstract}


\section{Introduction}
The attenuation of very high energy (VHE) \grays\ by pair production on the 
Galactic interstellar radiation field (ISRF) has previously been considered 
to be negligible \citep{Nikishov62,Protheroe1986}. 
The main contribution is thought to come from pair production on 
the cosmic microwave
background (CMB) where the effective threshold for attenuation is $\sim$100 TeV 
and a maximum is reached at about 2000 TeV, currently accessible only 
via air-shower experiments.
The Galactic ISRF photons are more energetic so that the effective 
threshold is lower ($\sim$100 GeV) and the attenuation increases slowly 
to a maximum around 100 TeV.
This covers the energy range of present day Imaging Atmospheric Cherenkov 
Telescopes, 
such as the HESS instrument.
A rough estimate of the attenuation of VHE \grays\ coming from the
Galactic center (GC), which uses a new ISRF \citep{PS05}, but assumes 
an isotropic angular distribution for the ISRF, shows that the effect is 
observable \citep{Z05}.
Given the essential
anisotropy of the Galactic ISRF with most of the photons going outwards from 
the inner Galaxy, the
effect depends on the position of the source of VHE photons and its orientation
relative to the observer in the Galactic plane. 
We calculate the attenuation of VHE \grays\ due to pair production 
with the Galactic photon field using the 
total ISRF over the entire Galaxy on a fine spatial grid 
which takes into account the nonuniform spatial 
and anisotropic angular distribution of background photons.

\vspace{2\baselineskip}
\section{Interstellar Radiation Field}
The essential ingredients to calculate the Galactic ISRF are a model for 
the distribution of stars in the Galaxy, a model for the dust distribution and
properties, and a treatment of scattering, absorption, and subsequent 
re-emission of the stellar light by the dust.
We briefly describe our ISRF calculation, which is a further development 
of the work reported by \cite{PS05}; full details will be given 
in a forthcoming paper (Porter \& Strong, in preparation).

Our stellar model assumes a type classification based on that used in the 
SKY model of \cite{Wainscoat1992}.
It includes 87 stellar classes encompassing main sequence stars, AGB stars 
and exotics.
For each stellar type there is a local star number density, 
scale height above the plane, 
fraction of local number density in each of several discrete spatial 
components, 
and spectrum in standard photometric filters.
The stars are distributed in seven geometrical 
components: thin and thick disc, halo, bulge, bar, ring, and spiral arms.
Spectra for normal stars are taken from the synthetic spectral 
library of \cite{Girardi2002}. 
Spectra for AGB stars and exotics are as given in the SKY model.

\begin{figure}[t]
\vspace{1\baselineskip}
\centerline{\includegraphics[width=3.5in]{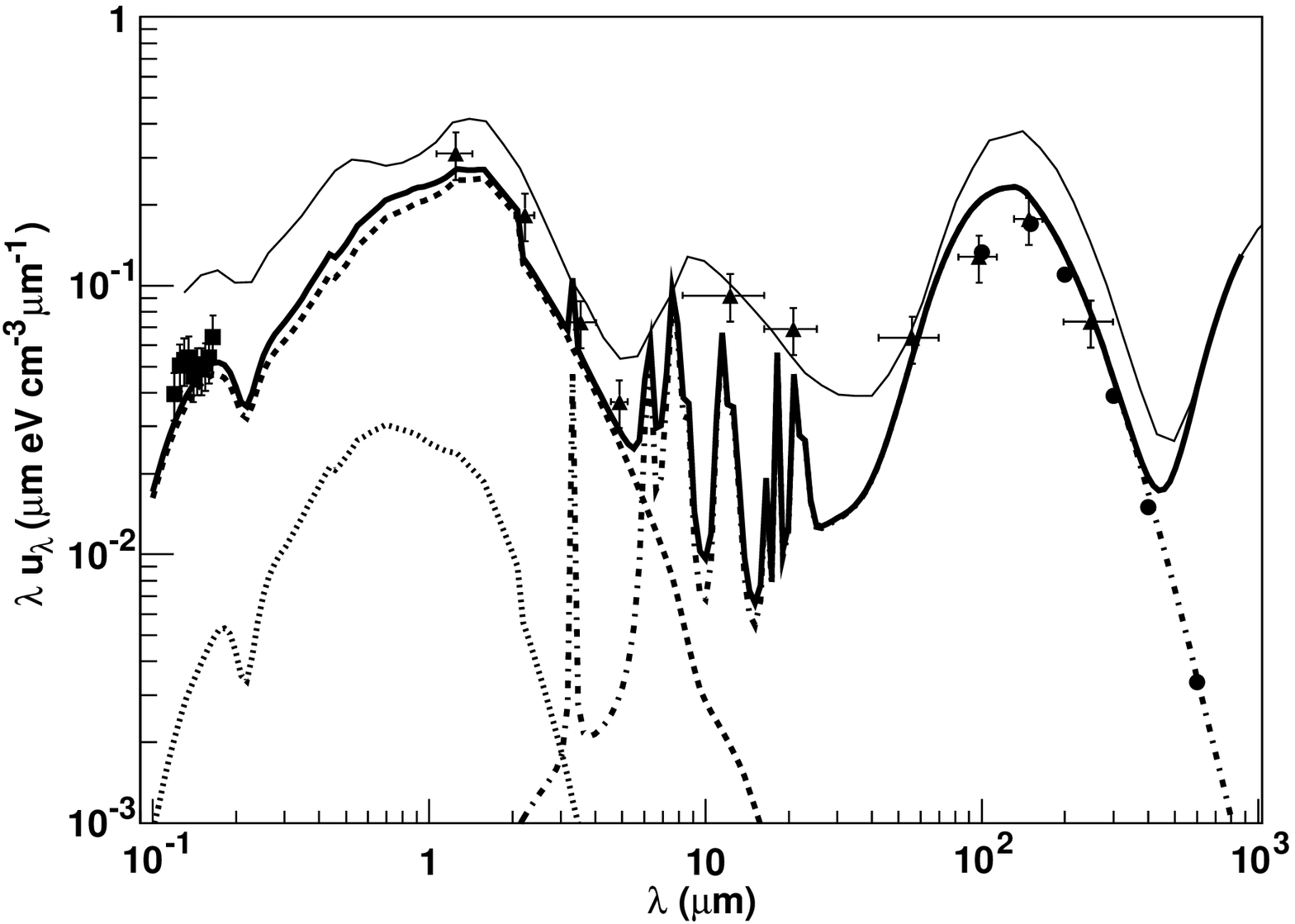}}\vspace{2\baselineskip}
\centerline{\includegraphics[width=3.5in]{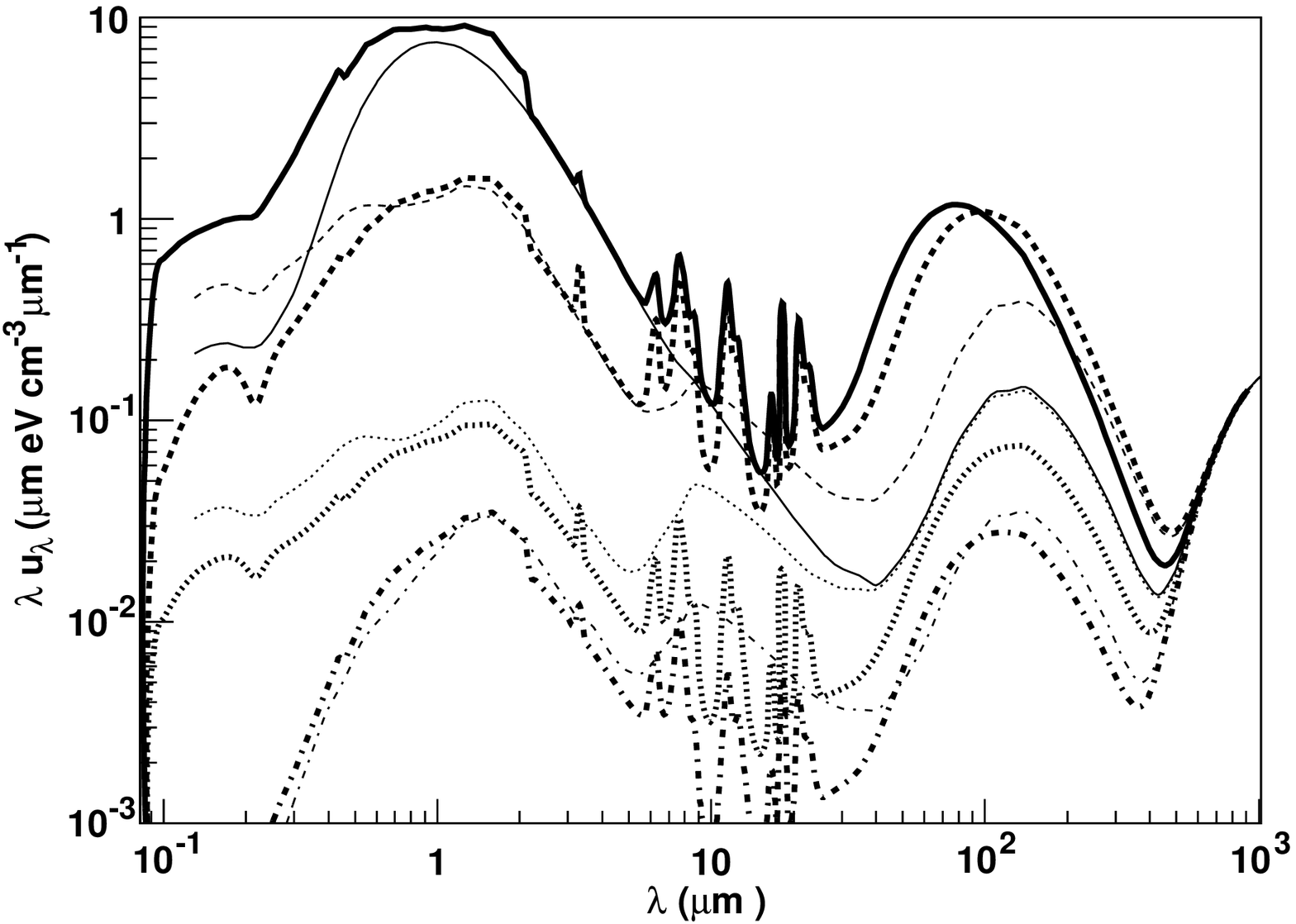}}
\vspace{1\baselineskip}
\caption{Interstellar radiation field energy density.
Local interstellar radiation field (upper): 
heavy solid line, total radiation field including CMB; 
heavy dashed line, contribution by stars; 
heavy dotted line, scattered light; 
heavy dot-dashed line, infra-red;
thin solid line, local ISRF from \cite{Strong2000}.
Data: squares, Apollo \citep{Henry1980}; 
triangles, DIRBE \citep{Arendt1998}; 
circles, FIRAS \citep{Finkbeiner1999}.
Interstellar total radiation field radial variation (lower): 
solid line, $(R, z) = (0$ kpc, 0 kpc);
dashed line, $(R, z) = (4$ kpc, 0 kpc);
dotted line, $(R, z) = (12$ kpc, 0 kpc);
dash-dotted line, $(R, z) = (16$ kpc, 0 kpc).
Heavy lines are for our ISRF; thin lines for the ISRF of \cite{Strong2000}.
\label{fig1}}
\vspace{1\baselineskip}
\end{figure}

We assume a dust model 
including graphite, polycyclic aromatic hydrocarbons (PAHs), and silicate.
Dust grains in the model are spherical and the absorption 
and scattering efficiencies for graphite, PAHs, and silicate grains 
are taken from \cite{Li2001}.
The grain model abundance and size distribution are taken from 
\cite{Weingartner2001} (their best fit Milky Way model), and 
a purely neutral interstellar medium is assumed.
We consider only coherent scattering, and a Henyey-Greenstein angular 
distribution function \citep{Henyey1941} is used in the scattering calculation.
The stochastic heating of grains smaller than $\sim0.1$ $\mu$m
is treated
using the ``thermal continuous'' approach of \cite{Draine2001}; 
we calculate the equilibrium heating of larger dust grains by balancing 
absorption with re-emission as described by \cite{Li2001}.

Dust is assumed to follow the Galactic gas distribution. 
We use the gas model
for neutral and molecular hydrogen given by \cite{Moskalenko2002}.
The radial variation in the Galactic metallicity gradient is taken to be 
0.07 dex/kpc \citep[][and references therein]{Strong2004b}.

A cylindrical geometry is adopted for the radiation field calculation.
Our calculations are simplified by assuming symmetry in azimuth and 
about the Galactic plane.
The maximum radial extent is taken to be $R_{\rm max} = 20$ kpc.
The maximum height above the Galactic plane is taken to be $z_{\rm max} = 5$ 
kpc. The Sun is located at $R_S = 8.5$ kpc from the GC.
The Galaxy is divided into elements of 
equal volume and the total radiation field is calculated for each.

The radiation field from stellar and scattered light is obtained using a
modified form of the so-called partial intensity method 
\citep{Baes2001}. 
Henceforth, 
we will refer to the total stellar and scattered light as the ``optical'' 
radiation field.
The infra-red radiation field is obtained by using the optical 
radiation field to calculate the
dust emissivity for stochastic and equilibrium heating.
The dust emission is integrated to obtain the infra-red 
radiation field throughout the Galaxy.

Fig.~\ref{fig1} (upper panel) 
shows our calculated local ISRF including the CMB; 
also shown in the figure is the local ISRF calculated by 
\cite{Strong2000}.
The agreement of our computed ISRF with the data is generally good. 
We note that our new ISRF provides an improved fit to the observations 
around 100 $\mu$m compared to the earlier model of \cite{Strong2000}.
This is particularly important for our optical depth calculation, since the 
majority of the attenuation is on these more numerous less-energetic photons
of the ISRF.

Fig.~\ref{fig1} (lower panel) shows the radial variation in the Galactic 
plane of our ISRF (thick lines) together with that of \cite{Strong2000} 
(thin lines).
Toward the inner Galaxy, our ISRF predicts a significantly higher energy 
density than previously described, particularly for the infra-red component.
This arises because of the coupling between the optical radiation field 
and the infra-red emission: in our calculation, 
the optical radiation field is used as 
direct input to the dust heating calculation, which was not
done in previous work.
The intense optical radiation field toward the inner Galaxy heats the 
dust to warmer temperatures, increasing the emission in the infra-red.
Even though the optical emission does not cause a direct enhancement to 
the attenuation at GeV to TeV energies -- the energy of these photons
is typically a few eV, lowering the threshold, however 
their number density is too low to provide significant absorption -- 
it is essential to calculate this component of the ISRF  
to obtain the correct emission and angular distribution for the infra-red. 
Further discussion of the new ISRF is deferred to a forthcoming paper (Porter
\& Strong, in preparation).

\begin{figure}[t]
\centerline{\includegraphics[width=3.5in]{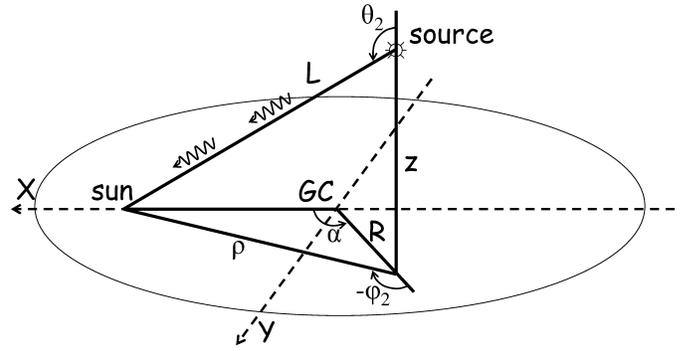}}
\caption{Angles involved in the calculation: $(R,z,\alpha)$, galactocentric
coordinates of the source of VHE photons; 
$(\theta_2,\phi_2)$, angles of the VHE photon; 
GC marks the Galactic center; and the observer's position is marked by ``sun''.
\vspace{1\baselineskip}
\label{fig2}}
\end{figure}

\section{Calculations}
The optical depth for VHE \grays\ is given by the general formula:
\begin{eqnarray}
\tau_{\gamma\gamma}(E)&&=   \label{tau}\\
&&\int_L dx \int d\varepsilon \int d\Omega_1 \,
\frac{dN(\varepsilon,\Omega_1,x)}{d\varepsilon d\Omega_1}\sigma_{\gamma\gamma}
(\varepsilon_c) (1-\cos\theta),\nonumber
\end{eqnarray}
where $dN(\varepsilon,\Omega_1,x)/d\varepsilon d\Omega_1$ is the 
differential number density of background photons at the 
point $x$, $\varepsilon$ is the background photon energy, 
$d\Omega_1=d\cos\theta_1 d\phi_1$ is a solid angle, 
$\sigma_{\gamma\gamma}$ is the total cross section for the pair production 
process $\gamma\gamma\to e^+e^-$ \citep{JR80},
$\varepsilon_c=[\frac12\varepsilon E (1-\cos\theta)]^{1/2}$
is the center-of-momentum system energy of a photon, 
and $\theta$ is the angle between the 
momenta of the two photons in the observer's frame. 
The integral over $x$ should be taken along the path of the \grays\ 
from the source to the observer. 

The ISRF is cylindrically symmetric so that the photon angular distribution
depends on $R$ and $z$ only.
In Fig.~\ref{fig2} we illustrate the galactocentric coordinate system 
$(R,z,\alpha)$. 
To calculate $\cos\theta$, the polar and azimuthal angles of the VHE photon, 
$\theta_2$ and $\phi_2$, are derived:
\begin{eqnarray}
&& \rho^2 = R^2+R_s^2-2RR_s\cos\alpha,\nonumber\\
&& \left\{ \begin{array}{l}
\sin\phi_2  =  -(R_s/\rho) \sin\alpha,\smallskip\\
\cos\phi_2  =  -(\rho^2+R^2-R_s^2)/2R\rho,
\end{array} \right. \nonumber\\
&& \left\{ \begin{array}{l}
\sin\theta_2 = \rho\, (\rho^2+z^2)^{-1/2},\smallskip\\
\cos\theta_2 = -(1-\sin^2\theta_2)^{1/2},
\end{array} \right. \\
&& \cos\theta =\cos\theta_1\cos\theta_2+\sin\theta_1\sin\theta_2\cos(\phi_1-\phi_2),\nonumber
\label{angles}
\end{eqnarray}
where $R_s$ is the galactocentric radius of the Sun.
The integration of eq.~(\ref{tau}) is done numerically.

For the calculation of the optical depth in the CMB field we use the formula:
\begin{eqnarray}
\tau_{\gamma\gamma}^{CMB}(E) = &&\\
\frac{-4kT}{(\hbar c)^3 \pi^2 E^2} &&
\int_L dx \int_{m_ec^2}^{\infty} d\varepsilon_c \varepsilon_c^3 
\sigma_{\gamma\gamma}
(\varepsilon_c) \log\left( 1-e^{-\varepsilon_c^2/E kT}\right),\nonumber
\label{tau_cmb}
\end{eqnarray}
where $kT$ is the CMB temperature, and $m_ec^2$ is the electron rest mass.

\section{Results}
\begin{figure}[t]
\centerline{\includegraphics[width=3.5in]{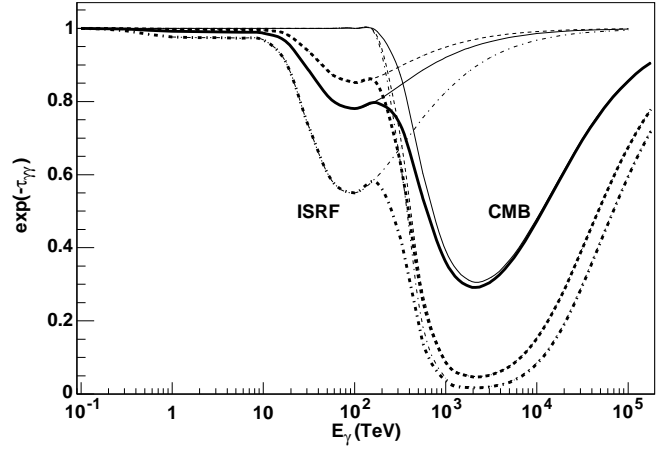}}
\vspace{1\baselineskip}
\caption{Transmittance of VHE \grays\ as a function of \gray\ energy. 
Solid line: $(R, z, \alpha) = (0$ kpc, 0 kpc, $0^\circ)$ -- $L = 8.5$ kpc;
Dashed line: $(R, z, \alpha) = (20$ kpc, 0 kpc, $90^\circ)$ -- $L = 21.8$ kpc;
Dash-dotted line: $(R, z, \alpha) = (20$ kpc, 0 kpc, $180^\circ)$ -- $L = 28.5$
kpc.
Thick lines give the total transmittance curve including the ISRF and CMB.
Left-most thin lines give the transmittance for the ISRF only; right-most 
thin lines for the CMB only.
\vspace{1\baselineskip}
\label{fig3}}
\end{figure}

Fig.~\ref{fig3} shows the attenuation for selected positions 
$(R, z, \alpha)$ as a function of incident \gray\ energy.
For sources located at the GC the attenuation is 
$\sim$12\% at 30 TeV and $\sim$23\% at 100 TeV.
In Table~\ref{TabTau} we give our optical depth results
for selected values of $R$ and $z$ for 
$\alpha = 0^\circ, 90^\circ$ and $180^\circ$ 
at 30 and 100 TeV without contribution by the CMB.
The attenuation is strongest for VHE \gray\ sources located toward the inner 
Galaxy and on its farside. 

\begin{figure*}[t]
\centerline{
\includegraphics[width=3.5in]{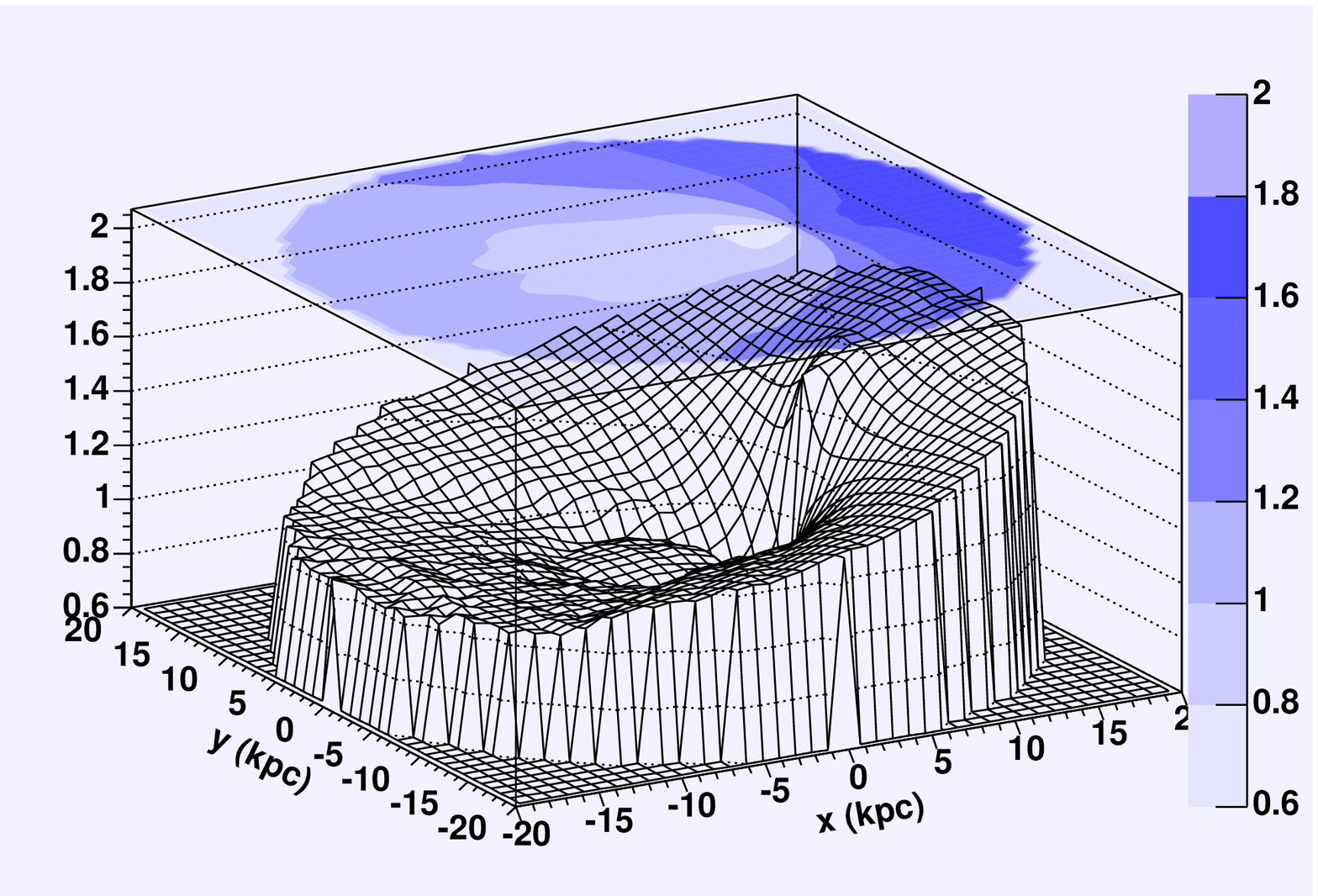}
\includegraphics[width=3.5in]{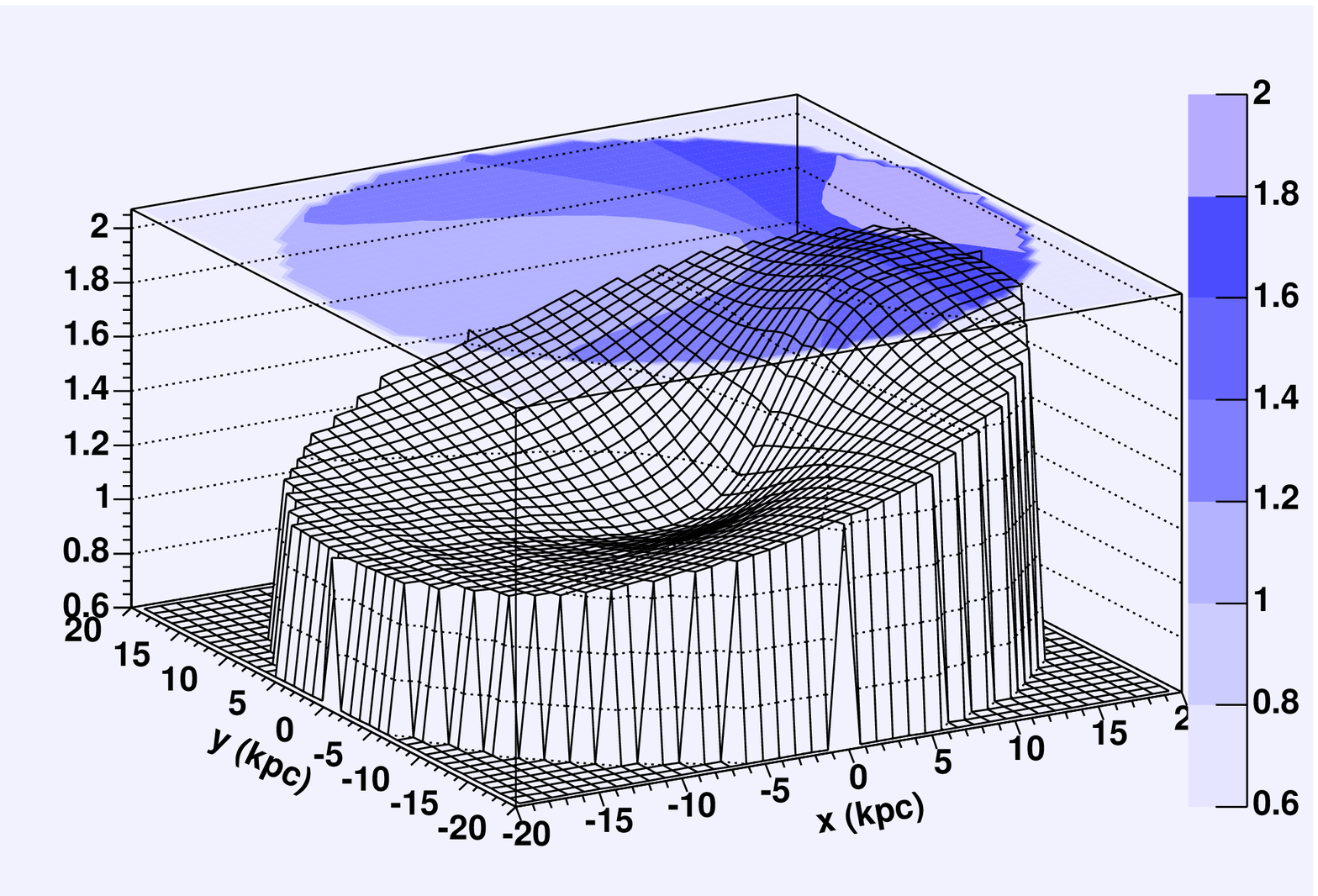}}
\vspace{1\baselineskip}
\caption{Effect of anisotropic ISRF on optical depth calculation at 100 TeV.
Shown is the ratio of the optical depths 
$\tau_{\gamma\gamma}/\tau_{\gamma\gamma}^{\rm iso}$ as a function of position 
$(x,y)$ for a source located at $z = 0$ kpc (left) and $z = 5$ kpc (right).
The solar system is located at $(x, y)$ = (8.5 kpc, 0 kpc), exactly where the 
optical depth ratio rises sharply.
\label{fig4}}
\vspace{1\baselineskip}
\end{figure*}

To illustrate the effect of the anisotropic radiation field on the attenuation
calculation, we show in Fig.~\ref{fig4} the ratio of the optical depths 
$\tau_{\gamma\gamma}/\tau_{\gamma\gamma}^{\rm iso}$, where $\tau_{\gamma\gamma}$ is 
calculated using the full angular distribution of the ISRF and 
$\tau_{\gamma\gamma} ^{\rm iso}$ is calculated assuming an isotropic distribution,
for a source located 
at $z = 0$ kpc (upper panel) and $z = 5$ kpc (lower panel) 
emitting 100 TeV \grays\ as a function of position.
The variation of $\tau_{\gamma\gamma}/\tau_{\gamma\gamma} ^{\rm iso}$ 
over the Galaxy as seen from Earth is non-trivial.
For sources located in the Galactic plane between the GC and solar system 
the ratio is less than one since the majority of the ISRF photons are 
coming from the GC direction, $\cos\theta \rightarrow 1$ in eq.~(\ref{tau}) 
and the interactions are mainly following, leading to a lower pair production
probability when the ISRF angular distribution is taken into account.
The reverse situation, $\tau_{\gamma\gamma}/\tau_{\gamma\gamma} ^{\rm iso} > 1$, 
occurs for \grays\ interacting with the ISRF in the outer Galaxy, where the 
majority of interactions are now head-on, $\cos\theta \rightarrow -1$ 
in eq.~(\ref{tau}). 
For sources located beyond the GC, the ratio is $\sim 1$. 
The interpretation of this case is straightforward: \grays\ emitted on
the farside of the Galaxy toward the solar system 
have mainly head-on absorption interactions until 
they reach the GC, whereupon the interactions become mainly following.
The isotropic case averages the angular distribution, and therefore leads 
to $\tau_{\gamma\gamma}/\tau_{\gamma\gamma} ^{\rm iso} \sim 1$.
We note that the optical depth ratio increases monotonically with $z$, and 
is a result of the progressively more head-on nature of the \gray\ absorption
interactions for sources located at larger distances from the plane.

\begin{deluxetable}{ccrrrrrrrr}
\tablecolumns{10} 
\tablewidth{0pc} 
\tabletypesize{\scriptsize}
\tablecaption{Optical depth $\tau_{\gamma \gamma}$ at 30 and 100 TeV} 
\tablehead{ 
\colhead{} & \colhead{} & \colhead{} &
\multicolumn{3}{c}{30 TeV} & \colhead{} & \multicolumn{3}{c}{100 TeV} \\
\cline{4-6} \cline{8-10} \\
\colhead{R, kpc} &
\colhead{z, kpc} & 
\colhead{} &
\colhead{$\alpha = 0^\circ$} & 
\colhead{$90 ^\circ$} & 
\colhead{$180^\circ$} &
\colhead{} &
\colhead{$\alpha = 0^\circ$} & 
\colhead{$90 ^\circ$} & 
\colhead{$180^\circ$} 
}
\startdata 
 0 & 0 & & 0.12 & \nodata & \nodata & & 0.25 & \nodata & \nodata\\ 
 5 & 0 & & 0.01 & 0.15 & 0.22 & & 0.05 & 0.32 & 0.41\\
10 & 0 & & 0.01 & 0.08 & 0.28 & & 0.02 & 0.21 & 0.54\\
15 & 0 & & 0.02 & 0.07 & 0.31 & & 0.05 & 0.18 & 0.58\\
20 & 0 & & 0.03 & 0.07 & 0.32 & & 0.06 & 0.16 & 0.60\medskip \\ 
 0 & 5 & & 0.03 & \nodata & \nodata & & 0.09 & \nodata & \nodata\\ 
 5 & 5 & & 0.01 & 0.04 & 0.06 & & 0.05 & 0.10 & 0.15\\
10 & 5 & & 0.02 & 0.05 & 0.09 & & 0.04 & 0.12 & 0.21\\
15 & 5 & & 0.03 & 0.05 & 0.12 & & 0.05 & 0.13 & 0.26\\
20 & 5 & & 0.03 & 0.06 & 0.14 & & 0.06 & 0.13 & 0.29\\ 
\enddata 
\label{TabTau}
\end{deluxetable} 

Our results show that the attenuation of VHE \grays\ by the Galactic
radiation field may be marginally observable by the HESS instrument
which has an effective sensitivity up to several tens of 
TeV \citep{Aharonian2005}.
The attenuation of VHE \grays\ from the sources on the Galaxy's farside 
will essentially steepen their spectra above $\sim$10 TeV.
In any case correction of source spectra for absorption is required. 
Interestingly, observations by future high energy experiments of the 
steepening of the spectra of 
Galactic sources in the GC region and beyond may serve as a probe 
of the Galactic ISRF.

\acknowledgments

I.\ V.\ M.\ acknowledges partial support from NASA
Astronomy and Physics Research and Analysis Program (APRA) grant.
T.\ A.\ P.\ acknowledges partial support from the US Department of Energy.

\end{document}